\documentstyle[12pt,aasms4]{article}

\def\kms{km s$^{-1}$\ }

\received{}
\accepted{}
\journalid{}{}
\articleid{}{}
\lefthead{Fisher et al.}
\righthead{Spectra of SN 1991T}

\begin{document}

\title{On the Spectrum and Nature of the Peculiar Type~Ia Supernova
1991T}

\author{ADAM FISHER, DAVID BRANCH, KAZUHITO HATANO, and E. BARON}

\affil{Department of Physics and Astronomy, 
University of Oklahoma, Norman, OK 73019--0225, USA}

\begin{abstract}

A parameterized supernova synthetic--spectrum code is used to study
line identifications in the photospheric--phase spectra of the
peculiar Type~Ia SN~1991T, and to extract some constraints on the
composition structure of the ejected matter.  The inferred composition
structure is not like that of any hydrodynamical model for Type~Ia
supernovae.  Evidence that SN~1991T was overluminous for an SN~Ia is
presented, and it is suggested that this peculiar event probably was a
substantially super--Chandrasekhar explosion that resulted from the
merger of two white dwarfs.

\end{abstract}

 
\section{Introduction}

SN~1991T was a well observed and spectroscopically peculiar Type~Ia
supernova (Filippenko et~al. 1992; Ruiz--Lapuente et~al. 1992;
Phillips et~al. 1992; Branch, Fisher, \& Nugent 1993).  Before and
around the time of maximum light its optical spectrum showed strong
lines of Fe~III rather than the usual SN~Ia lines of singly ionized
elements of intermediate mass, and although the deep red Si~II
absorption that is characteristic of SNe~Ia finally did develop after
maximum light, it never reached its usual strength.
 
In this paper we report the results of a study of photospheric--phase
spectra of SN~1991T using the parameterized supernova
spectrum--synthesis code SYNOW (Fisher et~al. 1997; Fisher 1998).  In
Section~2, previous studies of the optical spectra of SN~1991T are
briefly summarized.  Our method of ``direct'' spectral analysis is
described in Section~3, and results are presented in Section~4.  In
Section~5 we discuss evidence that SN~1991T was too luminous to be a
Chandrasekhar--mass explosion, and suggest that it probably was a
substantially super--Chandraskhar explosion resulting from the merger
of two white dwarfs.  A general discussion appears in Section 6.

\section{Previous Studies of SN~1991T Spectra}

Filippenko et~al (1992) presented optical spectra obtained from $-12$
to $+47$ days. (Thoughout this paper, epochs are in days with respect
to the date of maximum light, 1991 April 28; Lira et~al. 1998).  They
showed that although the premaximum optical spectra did not resemble
those of any other supernova, beginning near maximum light the usual
SN~Ia lines of intermediate--mass elements slowly developed, and
months after explosion the iron--dominated spectrum appeared almost
identical to that of a typical SN~Ia.  They identified the two strong
features in the premaximum optical spectrum with the Fe~III
$\lambda4404$ and $\lambda5129$ multiplets and concluded that the
composition of the outer layers was dominated by iron--group elements.
Given their inferred composition structure of a thin layer of
intermediate--mass elements sandwiched between inner and outer regions
dominated by iron--peak elements, they favored a double--detonation
model (the nearly complete incineration of a mildly
sub--Chandrasekhar--mass white dwarf by detonation waves propagating
inward and outward from the base of an accumulated helium layer) for
the origin of SN~1991T.  They noted that it was odd, then, that in
their $+6$ day spectrum an absorption near 7550\AA\ that is ordinarily
attributed to O~I $\lambda7773$ seemed to be present at its usual
strength.

Ruiz--Lapuente et~al (1992) presented optical spectra obtained on
seven consecutive nights from $-13$ to $-7$ days.  They too identified
the Fe~III lines, as well as lines of Ni~III, and they too concluded
that the outer layers had undergone complete burning to iron--peak
elements.  They supported that conclusion by presenting synthetic
spectra based on approximate NLTE calculations, for the
time--dependent nickel--cobalt--iron composition that results from the
radioactive decay of initially pure $^{56}$Ni.  As they noted, though,
the lines in their synthetic spectra tended to be stronger than the
observed lines.  Their spectra did not extend late enough in time for
them to encounter the oxygen line.  They did, however, discuss another
puzzle.  Their spectra showed no evidence for the high velocities of
the outer layers that would follow from complete burning to iron--peak
elements.  They suggested that the line--forming layers had been
decelerated upon encountering a low--density carbon--oxygen envelope
associated with a merger of two white dwarfs.

Phillips et~al. (1992) presented optical spectra obtained from $-13$
to $+66$ days.  From the appearance of the spectra they concluded that
the abundances of silicon, sulfur, and calcium in the outer layers
were unusually low, but they did not specify what was present in their
place.

Jeffery et~al. (1992) carried out parameterized synthetic--spectrum
calculations, based on the Sobolev approximation, an approximate
radiative--equilibrium temperature distribution, and full LTE,
including a Planckian rather than a resonance scattering source
function.  They modified the composition structure of the
carbon--deflagration model W7 (Nomoto, Thielemann, \& Yokoi 1984;
Thielemann, Nomoto, \& Yokoi 1986), by trial and error, to get
reasonable fits to observed optical (and some IUE near--ultraviolet)
spectra obtained at $-9$, $-3$, and $+10$ days.  The adopted
composition (displayed in their Figure~7) was complex, and not simply
dominated in the outer layers by iron--peak elements.  It was like
that of model W7 for ejection velocities less than 8500 \kms; heavily
modified between 8500 and 14,400 \kms in favor of iron--peak elements
at the expense of intermediate--mass elements; heavily modified
between 14,400 and 21,000 \kms in favor of both intermediate--mass
and iron--peak elements at the expense of carbon; and primarily carbon
and oxygen above 21,000 \kms.  Noting that somewhat weaker Fe~III
lines might have gone unrecognized in previous studies of other
SNe~Ia, they suggested that although the composition structure was
unusual the explosion history of SN~1991T probably was not
fundamentally different from that of a normal SN~Ia.

Mazzali, Danziger, \& Turatto (1995) also carried out parameterized
spectrum calculations, using a Monte Carlo code and a resonance
scattering source function.  They were restricted to using a
homogeneous composition above the photosphere at each of the seven
epochs considered.  For each epoch they used a combination of a mixed
W7 composition and the time--dependent $^{56}$Ni--decay composition.
They concluded that an unusually high temperature was partly
responsible for the weakness of the lines of singly ionized
intermediate--mass elements, but also that iron--peak elements
dominated the composition above 13,000 \kms. They favored a
``late--detonation'' (Yamaoka et~al. 1992) explosion mechanism in a
Chandrasekhar--mass white dwarf for the origin of SN~1991T.  (Mazzali
et~al referred to it as a ``delayed detonation'', but by custom that
term is used to refer a class of models that, unlike the late
detonations, do not synthesize $^{56}$Ni in the outer layers; e.g.,
Khokhlov, M\"uller, \& H\"oflich 1993).

Nugent et~al. (1995) calculated detailed NLTE spectra using the
PHOENIX code [cf. Hauschildt and Baron 1998, and references therein], 
for a fixed composition (W7 mixed 
above 8000 \kms, with titaniium enhanced by a factor of 10) and a 
series of temperatures.  They found that their series of synthetic
spectra gave a good account of many of the spectral differences among
SNe~Ia, from the peculiar, cool, ``weak'' SN~1991bg, through the
normal SNe~Ia, to the peculiar, warm, ``strong'' SN~1991T.  This
showed that to a certain extent the differences between the spectra of
SNe~Ia are due to temperature differences, and confirmed that in
SN~1991T a high temperature can enhance the Fe~III and Si~III lines
and weaken the lines of singly ionized elements. Still, the
temperature differences among SNe~Ia presumably are caused by
differences in the amounts of ejected $^{56}$Ni, and differences in
the composition structures certainly are to be expected.

\section{Spectrum Synthesis Procedure}

In this paper we use the fast, parameterized, supernova
spectrum--synthesis code SYNOW to make a ``direct'' analysis (Fisher
et~al. 1997) of spectra of SN~1991T.  The goal is to study line
identifications and determine intervals of ejection velocity within
which the presence of lines of various ions are detected, without
initially adopting any particular composition structure.  The
composition and velocity constraints that we obtain with SYNOW then
can provide guidance to those who compute hydrodynamical explosion
models and to those who carry out computationally intensive NLTE
spectrum modeling.  The SYNOW code was described briefly by Fisher
et~al. (1997) and in detail by Fisher (1998).  In our work on SN~1991T
we have made extensive use of the paper by Hatano et~al. (1998), which
presented plots of LTE Sobolev line optical depths versus temperature
for six different compositions that might be expected to be
encountered in supernovae, and which presented SYNOW optical spectra
for 45 individual ions that can be regarded as candidates for
producing identifiable spectral features in supernova spectra.

Fisher et~al. (1997) concentrated on a high quality spectrum of the
normal Type~Ia SN~1990N that was obtained by Leibundgut~et~al. (1991)
at $-14$ days.  Fisher et~al. suggested that at this very early phase
an absorption feature observed near 6040\AA, which previously had been
attributed to moderately blueshifted Si~II $\lambda6355$, actually was
produced by highly blueshifted ($v > 26,000$ \kms) C~II $\lambda6580$,
indicating the presence of an outer high--velocity carbon--rich layer
in SN 1990N.  In this paper we suggest that SN 1991T also contained an
outer carbon--rich layer, but extending to lower velocity than in
SN~1990N.

We have studied spectra of SN~1991T ranging from $-13$ days to $+59$
days.  For comparison with each observed spectrum we have calculated a
large number of synthetic spectra with various values of the fitting
parameters.  These include: $T_{bb}$, the temperature of the
underlying blackbody continuum; $T_{exc}$, the excitation temperature;
$v_{phot}$, the velocity of matter at the photosphere; and $v_{max}$,
the maximum ejection velocity. For each ion that is introduced, the
optical depth of a reference line also is a fitting parameter, with
the optical depths of the other lines of the ion are calculated assuming 
Boltzmann excitation at $T_{exc}$.  In addition, we can introduce
restrictions on the velocity interval within which an ion is present;
when the minimum velocity assigned to an ion is greater than the
velocity at the photosphere, the line is said to be detached from the
photosphere.  The radial dependence of the line optical depths is
taken to be exponential with e--folding velocity $v_e = 3000$ km
s$^{-1}$ (with one exception to be discussed in section 4.2) and the
line source function is taken to be that of resonance scattering.  The
most interesting fitting parameters are $v_{phot}$, which as expected, 
decreases with time, and the individual ion velocity
restrictions, which constrain the composition structure.

\section{Spectrum Synthesis Results}

In this section we present comparisons of synthetic spectra with
observed spectra at several selected epochs, to illustrate how we
reach our conclusions about the composition structure of the ejecta.
The premaximum and postmaximum spectra are discussed separately,
because the the former were so peculiar while the latter became
increasingly normal.

\subsection{Premaximum}

In the premaximum spectra only two features have obvious
identifications: two strong features produced by the Fe~III
$\lambda4404$ and $\lambda5129$ multiplets.  Figure~1 compares a
$-13$ day observed spectrum to a synthetic spectrum that has
$v_{phot}=15,000$ \kms and $T_{exc}=13,000$ K.  The optical depth of
the Fe~III reference line and the velocity at the photosphere have
been chosen such that the $\lambda5129$ multiplet gives a reasonable
fit to the strong P Cygni feature near 5000\AA.  Similarly, the
optical depth of the reference line of Ni~III has been chosen to give
a reasonable fit to the weaker absorption near 5300\AA; then, in the
synthetic spectrum, Ni~III also is responsible for the weak absorption
near 4700\AA.  We consider the presence of Ni~III lines in the
observed spectrum to be probable. 

The diagnostic value of a spectral feature is not simply proportional
to its strength.  A weak feature in the premaximum spectra that is of
special interest to us is the broad, shallow absorption near 6300\AA,
which definitely is a real feature because it can be seen in almost
all of the spectra obtained by Filippenko et al. (1992),
Ruiz--Lapuente et~al. (1992) and Phillips et al. (1992) earlier than
$-6$ days; this feature was not mentioned in the previous discussions
of the SN~1991T spectra.  As shown in Figure~1, C II $\lambda6580$ can
account for this feature, and in spite of searching for a plausible
alternative with the help of the plots of Hatano et~al. (1998), we
have none to offer.

Figure~2 compares a $-4$ day observed spectrum to a synthetic
spectrum that has $v_{phot}=10,000$ \kms and $T_{exc}=13,000$ K.  The
two strong Fe~III features are fit reasonably well, and Ni~III and
Ni~II probably account for some of the other observed features.
Additional ions of iron--peak elements would need to be introduced to
account for the observed flux deficiency at $\lambda < 3500$ \AA.  The
narrow absorption near 4400\AA\ is well fit by the Si~III
$\lambda4560$ multiplet, which is the strongest optical multiplet of
Si~III and was identified in SN~1991T by Jeffery et al. (1992).  Note
that at this phase, there is a weak absorption near 6200\AA, rather
than the 6300\AA\ of earlier phases.  In general, supernova absorption
features shift redward as the photosphere recedes with respect to the
matter, or they remain unshifted if the line has become detached.  A
blueward shift of an absorption is almost a sure sign that a new line
is beginning to make a significant contribution.  In this case the new
line is very likely Si~II $\lambda6355$, which in normal SNe~Ia is
very strong at this phase.  (In the synthetic spectrum, lines of Ni~II
and Fe~III also are affecting this feature.)

From the premaximum spectra we infer that iron, silicon, and probably
nickel were present above 10,000 \kms, with iron being detected up to
20,000 \kms.  If the nickel identifications are correct, then freshly
synthesized iron--peak elements were present in these outer layers.
Carbon seems to have been present down to at least 15,000 \kms.
Between $-13$ and $-4$ days, the red absorption feature apparently
transformed from mainly C~II forming above about 15,000 \kms to mainly
Si~II forming above 10,000 \kms.  A very careful study of a good
series of spectra obtained within this time interval might better
determine the minimum velocity at which C~II was present.  If the C~II
line became detached before the Si~II line made its appearance, the observed
absorption minimum whould have remained constant at the detachment
velocity until the development of the Si~II line started to cause a
shift to the blue.

\subsection{Postmaximum}

Figure~3 shows a $+6$ day observed spectrum.  By this time the
spectrum had begun to look much less peculiar.  The synthetic spectrum
has $v_{phot}=9500$ \kms and $T_{exc}=10,000$ K.  The synthetic
feature near 5000\AA\ is now a blend of Fe~III and Fe~II lines.  The
observed S~II and Si~II features are weaker and narrower than in
normal SNe~Ia.  In the synthetic spectrum of the top panel, a maximum
velocity of 12,000 \kms for S~II lines has been introduced to fit the
absorptions near 5300 and 5500\AA, and a maximum velocity of 15,000
\kms has been used to fit the Si~II absorption near 6200 \AA.  The
lower panel shows how the fit degrades when these maximum velocities
are not used.

The observed spectrum that appears in Figures~4 and 5 was obtained by
Meikle et~al. (unpublished) at $+59$ days.  The synthetic spectra have
$v_{phot}=4000$ \kms and $T_{exc}=10,000$ K.  The upper panel of
Figure~4 shows that resonance scattering by permitted lines of just
three ions --- Fe~II, Ca~II, and Na~I --- can give a reasonable
account of most of the features in the observed spectrum, although the
height of the synthetic peaks in the blue indicate that the synthetic
spectrum is underblanketed.  In the synthetic spectrum Na~I and Ca~II are 
detached at 9000 \kms, and an abrupt decrease in the Fe II line
optical depths, by a factor of 10, has been introduced at 10,000
\kms.  This is inferred to be a measure of the maximum velocity of
the iron--peak core. The structure of the synthetic spectrum near
5000\AA\ is quite sensitive to the velocity of the Fe~II line optical
depth discontinuity (Fisher 1998).

The major shortcoming in the upper panel of Figure~4 is that the broad
minimum observed near 7000\AA\ is not reproduced by the synthetic
spectrum.  (The weak synthetic absorptions in the vicinity are from
Fe~II.)  A possible identification for this feature is [O~II]
$\lambda\lambda7320,7330$ (Fisher~1998, Hatano et~al. 1998).  For a
forbidden transition the natural first approximation to the source
function would be the Planck function evaluated at the local electron
temperature, but instead of introducing a whole new fitting function
involving the radial dependence of the electron temperature, we have
simply retained the resonance scattering source function.  In the
lower panel of Figure~4 the [O~II] feature is calculated with a line
optical depth that is detached at 10,000 \kms, where $\tau=0.5$, and
has a shallow radial gradient ($v_e=20,000$ \kms) such that
$\tau=0.24$ at the maximum [O~II] velocity of 25,000 \kms. The upper
panel of Figure~5, which is like the upper panel of Figure~4 but with
the [O~II] feature included in the synthetic spectrum, looks better
than the upper panel of Figure~4.  The lower panel of Figure~5 shows
how the fit degrades when Na~I is not detached, the discontinuity in
the Fe~II line optical depths is not introduced, and the upper and
lower velocity limits on [O~II] are dropped.

 From a spectroscopic point of view the [O~II] identification is
attractive and plausible.  Moreover, Kirshner et~al. (1993) discussed
the velocity interval in which the O~I $\lambda7773$ line could be
detected in the early spectra of a small sample of well observed
SNe~Ia.  For SN~1991T they estimated that the O~I line had a
significant optical depth from at least as low as 9000 \kms to at
least as high as 19,000 \kms, which is not inconsistent with what we
are using for [O~II].  But the mass and associated kinetic energy of
the oxygen that would be required to produce a significant [O~II] line
optical depth, at such high velocities and at this fairly late phase,
appear to be high.  For this feature, with its very low transition
probability, just producing a uniform line optical depth of 0.2
requires

$$ M=0.26 v_{25}^3 t_{77}^2 f_O^{-1} M_\odot;\ \ E_K=10^{51} v_{25}^5
t_{77}^2 f_O^{-1} \rm erg $$

\noindent where $v_{25}$ is the maximum velocity in units of 25,000
\kms, $t_{77}$ is the time since explosion in units of 77 days
(allowing for a rise time to maximum of 18 days), and $f_O$ is the
fraction of all oxygen that is in the lower level of the transition.
It is possible that essentially all of the oxygen is singly ionized at
this phase, but $f_O$ must be significantly less than unity
because the lower level of the transition is 3.3~eV above the ground
level of singly ionized oxygen. (The ground level of the transition
is, at least, the metastable lowest level of the doublets, the true
ground level of singly ionized oxygen being a quartet.)

We have no plausible alternative to offer for the 7000\AA\ minumum.
If it is not an absorption feature at all, then the continuum must be
at a level that is considerably lower than we have adopted to obtain
the fits shown in the top panels of Figures~4 and 5.  We suspect that
the [O~II] identification is correct, but in view of the mass and
energy problem the maximum [O~II] velocity may need to be somewhat
lower than the 25,000 \kms that we have used.

\subsection{Summary of the Inferred Composition Structure}

In Paper~I we presented evidence for C~II in SN~1990N at $v > 26,000$
km s$^{-1}$.  It is thought that when SNe~Ia are arranged in a
sequence from powerful events like SN~1991T to weak ones like
SN~1991bg, SN~1990N belongs on the powerful side (e.g., Phillips 1992;
Nugent et~al. 1995).  One might then expect events that are weaker
than SN 1990N to have unburned carbon extending down to velocities
lower than 26,000 km s$^{-1}$, and our SYNOW studies of other SNe~Ia
suggest that this generally is the case (Fisher 1998).  Similarly, one
might expect events like SN~1991T that are thought to be stronger than
SN~1990N to have a minimum velocity of carbon that is higher than
26,000 \kms.  But in SN~1991T we find evidence for C~II moving at
least as slow as about 15,000 km s$^{-1}$.  At the same time, we find
evidence that the iron--peak core of SN~1991T extended out to
velocities at least as high as in SN~1990N and other SNe Ia.
Independently, Mazzali et~al. (1998) list outer velocities of the iron
core for 14 SNe~Ia, inferred from nebular--phase spectra, and they
find that SN~1991T has the highest velocity in their sample.  Thus
SN~1991T seems to have had both slower unburned carbon and a faster
iron--peak core than SN~1990N, with its intermediate--mass elements
being confined to an unusually narrower velocity interval.

Figure~6 shows the velocity at the photosphere as a function of time.
The pause in the velocity decrease, around 10,000 \kms, presumably
reflects an increase in the density or the opacity near the outer edge
of the iron--peak core.  Our combined constraints on the composition
structure are shown in Figure~7. The inferred structure is not like
any hydrodynamical model that has been published.  A speculation about
the cause of the peculiar composition structure of SN~1991T will be
offered in Section~5, after the implication of the high luminosity of
SN~1991T is considered in the next section.

\section{The Luminosity of SN~1991T} 

NGC 4527, the parent galaxy of SN~1991T, is in the southern extension
of the Virgo cluster complex. According to the Nearby Galaxies
Catalogue (Tully 1988) it is a member of the same small group of
galaxies (group 11 --4 in Tully's notation) as NGC 4536 and NGC 4496,
the parent galaxies of the normal Type Ia SNe 1981B and 1960F.  These
three galaxies have similar heliocentric radial velocities of 1730,
1866, and 1738 km s$^{-1}$, respectively, and they are the three
brightest galaxies in the group.  On the sky, NGC 4527 is 1.9 degrees
from NGC 4496 and only 0.6 degrees from NGC 4536.  Peletier \& Willner
(1991) found nearly identical Tully--Fisher distances for NGC 4527 and
NGC 4536.  Independently, Pierce (1994) obtained nearly identical
Tully--Fisher distance moduli for all three of these galaxies. Tully
(1988), Tully, Shaya, and Pierce (1992), and Peletier \& Willner
(1991) all agreed that this galaxy group is on the near side of the
Virgo cluster complex.  Since then, Saha et al. (1996a) have
determined a Cepheid--based distance modulus of $\mu = 31.10 \pm 0.13$
for NGC 4536, and similarly Saha et al. (1996b) obtain $\mu = 31.03
\pm 0.14$ for NGC 4496. Pending a direct Cepheid--based determination
of the distance to NGC 4527, which is to be attempted (A.~ Saha 1998,
personal communication), we assume here that the distance modulus to
NGC 4527 is $\mu = 31.07 \pm 0.13$ ($D = 16.4 \pm 1.0$ Mpc).

The magnitudes and luminosities of SNe 1960F, 1981B, and 1991T are
compared in Table 1.  The $B$ and $V$ peak apparent magnitudes are
from Saha et al. (1996b), Schaefer (1995a), and Lira et al. (1998),
respectively.  The observed $B$ and $V$ magnitudes of SNe 1960F and
1991T were similar, while those of SN 1981B were about 0.5--0.6 mag
fainter.

The extinction of these three events by dust in our Galaxy should be
negligible (Burstein \& Heiles 1982), but there are reasons to think
that SN 1991T was significantly extinguished by dust in its parent
galaxy. (1) In projection, at least, the event occurred near a spiral
arm of low surface brightness in NGC 4527, an Sb galaxy that is very
dusty and has a high inclination of 74 degrees.  For a good photograph
that shows the location of SN 1991T in NGC 4527, see Figure~1 of
Schmidt et al. (1994). (2) Photometric and spectroscopic observations
of SN 1991T at an age of 2--3 years have been interpreted by Schmidt
et al. (1994) in terms of a light echo caused by dust in NGC 4527.
(3) Interstellar lines of Ca II (Meyer \& Roth 1991) and Na I (Smith
\& Wheeler 1991; Filippenko et al. 1992; Ruiz-Lapuente et al. 1992),
at the redshift of NGC 4527, were detected in the spectra of SN 1991T.
Values of the color excess that have been estimated on the basis of
the strengths of the Na I lines include $E(B-V)=0.34$ by Ruiz-Lapuente
et al. (1992) and 0.13--0.23 by Filippenko et al. (1992). Although
these estimates are recognized to be uncertain, some significant
amount of extinction is to be expected. (4) Mazzali et al. (1995) and
Nugent et al. (1995) concluded from its spectral features that SN
1991T was hotter than normal SNe Ia, yet some of the broad--band
colors of SN 1991T were observed to be redder than those of normal SNe
Ia.  Phillips et al. (1992) estimated $E(B-V)=0.13$ by assuming that
the intrinsic $B-V$ color at maximum light was like that of normal SNe
Ia.  The revised photometry of Lira et~al. (1998) makes SN~1991T even
redder than had been thought, with $B_{max}-V_{max}=0.19\pm0.03$.

For SN~1991T we adopt $E(B-V)=0.2 \pm 0.1$. For SN 1981B we use
$E(B-V)=0.10 \pm 0.05$ (M.~M.~Phillips 1995, personal communication),
and we assume that the extinction of SN 1960F was negligible (Schaefer
1996; Saha et al. 1996b).  It may be worth noting that these estimates
are in accord with the galaxy ``dustiness'' categories of van den
Bergh \& Pierce (1990), who put NGC 4496 in category 1 (``some dust
visible''), NGC 4536 in category 2 (``dust easily visible'') and NGC
4527 in category 3 (``galaxy appears {\sl very} dusty''); only 12 of
the 230 galaxies in their sample were assigned to the very dusty
category.  With our adopted extinction estimates, the extinction--free
apparent ($B^0$ and $V^0$) and absolute ($M_B^0$ and $M_V^0)$
magnitudes of the normal SNe 1960F and 1981B become similar, while SN
1991T becomes brighter by 0.7--0.8 mag (see Table 1).

For the bolometric correction, $M_{bol} - M_V$, of normal SNe Ia such
as 1981B and 1960F, we adopt $0.1 \pm 0.1$ (H\"oflich 1995; Nugent et
al. 1995; Branch et al. 1997).  Even before being corrected for
extinction, SN 1991T had a larger fraction of its energy in the near
ultraviolet than do normal SNe Ia (Nugent et al. 1995; Schaefer 1995b;
Branch et al. 1997), therefore a smaller fraction of its total flux
was emitted in the $B$ and $V$ bands and its $M_{bol} - M_V$ was more
negative than that of normal SNe~Ia.  For SN~1991T we adopt $M_{bol} -
M_V = -0.1 \pm 0.1$, which probably is conservative for the present
argument.  The bolometric absolute magnitude of SN 1991T then exceeds
that of SN 1981B by $0.95 \pm 0.37$ mag (where the uncertainty in the
{\sl difference} between the distance moduli of NGC 4527 and NGC 4536
has been neglected), and the luminosity of SN 1991T exceeds that of SN
1981B by a factor between 1.7 and 3.4, with the best estimate being a
factor of 2.4 (see Table 1).

The peak luminosity of a Type Ia supernova can be written (Arnett 1982, 
Branch 1992)

$$ L = \alpha R(t_r) M_{Ni} \eqno (1) $$

\noindent where $R$, the instantaneous radioactivity luminosity per
unit nickel mass at the time of maximum light, is a known function of
the rise time $t_r$, $M_{Ni}$ is the mass of ejected $^{56}$Ni, and
$\alpha$ is dimensionless and of order unity.  For normal SNe Ia, such
as SN 1981B, characteristic values of $M_{Ni} = 0.6 M_\odot$, $t_r =
18$ days, and $\alpha = 1.2$ (e.g., H\"oflich \& Khokhlov 1996; Branch
et~al. 1997) give $L = 2.03 \times 10^{43}$ ergs s$^{-1}$.  This
corresponds to a bolometric absolute magnitude $M_{bol}=-19.57$, a
little brighter than, but not inconsistent with, the value implied by
the Cepheid distance and the adopted extinction.

At first glance one might think that sufficient overluminosity of SN
1991T with respect to SN 1981B could be achieved with a Chandrasekhar
mass, just by allowing the nickel mass to approach the Chandrasekhar
mass in SN 1991T (i.e., $1.4/0.6=2.33$, and we have estimated that
SN~1991T was more luminous than SN~1981B by a factor of 2.4).
However, there are severe problems with this simple picture, in which
nearly all of the ejected mass of SN 1991T is initially in the form of
$^{56}$Ni.  (1) Spectral lines formed by elements other than nickel,
cobalt, and iron show that the initial composition of SN~1991T was not
just $^{56}$Ni. (2) No hydrodynamical models of Chandasekhar--mass
explosions produce just $^{56}$Ni.  Even the pure detonation model of
Khokhlov et al. (1993) produced only $0.92\ M_\odot$ of $^{56}$Ni,
with the rest of the mass being in the form of other iron--peak
isotopes.  (3) A rapid expansion of the ejecta caused by the high
kinetic energy per gram, together with the prompt escape of gamma rays
emitted by $^{56}$Ni in the outer layers, would make the light curve
too fast, and with gammas escaping the value of $\alpha$ would be low.
For example, for the detonation model of Khokhlov et al. (1993),
H\"oflich \& Khokhlov (1996) calculated $\alpha=0.76$ (in their
notation it is $Q$).  (5) The optical and gamma--ray luminosities
depend on distance in the same way.  From the optical brightness of SN
1991T and the lack of detection of gamma rays by Lichti et al. (1994)
and Leising et al. (1995), the latter authors conclude that ``some way
of producing optically brighter but gamma--ray fainter supernovae
[compared to SN Ia models in the literature] is required to explain SN
1991T".  A Chandrasekhar--mass explosion that contained nearly a
Chandrasekhar mass of $^{56}$Ni would have a {\sl low} ratio of
optical and gamma--ray luminosities.  

If SN 1991T is at the same distance as SNe 1981B and 1960F then it is
unlikely that its luminosity can be explained in terms of a
Chandrasekhar mass ejection.

\section{Discussion}

If the luminosity of SN~1991T was too high to be explained in terms of
a Chandrarsekhar mass ejection, the only recourse would seem to be to
appeal to the explosion of a super--Chandrasekhar product of the
merger of two white dwarfs.  The question of whether mergers of white
dwarfs actually can produce explosions is not yet settled (e.g.,
Mochkovitch, Guerrero, \& Segretain 1997).  An attractive recent
suggestion was that of Iben (1997), who noted that because tidal
torques will spin up the premerger white dwarfs to rotate in near
synchronism with the orbital motion, huge shear forces will arise at
the onset of the merger.  If both white dwarfs ignite prior to or
during the merger, owing to shear and tidal heating, or if the
ignition of one of the white dwarfs then provokes the ignition of the
other, this might be a way to get not only a super--Chandrasekhar mass
ejection, but even a super--Chandrasekhar mass of $^{56}$Ni if such
should prove to be required.  Getting a super--Chandrasekhar mass of
$^{56}$Ni out of a thermonuclear explosion was difficult to envision
before Iben's suggestion.  It must be noted, though, that in the first
calculations of tidal heating during a merger, it failed by a narrow
margin to cause carbon to ignite (Iben \& Tutukov 1998).

It is interesting that on the basis of their population--synthesis
studies, Tutukov \& Yungelson (1994) predict that for ``young''
mergers, those that occur within 300 Myr of star formation, the
average combined mass is substantially super--Chandrasekhar, typically
about $2.1\ M_\odot$ (see their Figure~4).  Recall that SN 1991T
appears to have occurred near a spiral arm.  Preliminary indications
are that the several other events resembling SN~1991T that have been
discovered in recent years also tend to be associated with star
forming regions and/or to be significantly extinguished by dust
(A.~V.~Filippenko 1998, personal communication; P.~Garnavich 1998,
personal communication).  SN~1991T--type events may be from a younger
population than most SNe~Ia.

It also is interesting that Ruiz--Lapuente et al. (1992) discuss the
possible detection of a narrow circumstellar line of O I $\lambda
8446$ in their earliest spectrum of SN 1991T. Searching for narrow
circumstellar lines of hydrogen, helium, carbon, or oxygen is one the
best ways to probe the composition of the donor star in the binary
progenitor system of a SN Ia (Branch et al. 1995).  No signs of
circumstellar interaction, and no clear detections of narrow
circumstellar lines of hydrogen or helium have been found in {\sl any}
SN~Ia.

The proposition that SN 1991T was the result of a super--Chandrasekhar
merger is consistent with the findings of H\"oflich \& Khokhlov
(1996), who calculated light curves for a variety of SN Ia
hydrodynamical models and found that the slow light curve of SN 1991T
was best fit by models of Khokhlov et~al. (1993) that
had a super--Chandrasekhar ejecta and/or substantial amounts of
unburned carbon and oxygen in their outermost layers.  The particular
models that H\"oflich \& Khkokhlov cited as best fitting the light
curve have a peak $M_V \simeq -19.4$, which is not luminous enough for
SN 1991T.  However, in their models the underlying explosion was the
detonation of only 1.2 $M_{\odot}$, inside low--density carbon--oxygen
envelopes of 0.2, 0.4, and 0.6 $M_{\odot}$.  If the central explosion
had been the detonation of a Chandrasekhar mass the peak would have
been brighter.

In a very general sense, the composition structure expected of a
merger explosion might resemble the composition structure that has
been inferred for SN~1991T (cf. Ruiz--Lapuente et~al. 1992). An
unusually strong explosion produces a high--mass, high--velocity
iron--peak core surrounded by an unusually small mass of
intermediate--mass elements; this encounters a surrounding
low--density mass of carbon and oxygen which decelerates the
intermediate--mass elements and forces them into a narrow velocity
interval.  The super--Chandraskhar models of Khokhlov et~al. (1993)
had this sort of composition structure, but the minimum velocity of
unburned carbon was lower than 10,000 \kms, a value which is lower than we
infer from the spectra.  In this respect, too, replacing the 1.2
$M_\odot$ core with a Chandrasekhar mass would be a step in the right
direction.  None of this explains the presence of iron-peak elements
at higher velocity than the intermediate--mass elements, or the
possible coexistence in velocity space of iron--peak elements and
unburned carbon.  Eventually, only multidimensional hydrodynamical
studies of the merger process can tell us whether this is possible.

It should be noted that Liu, Jeffery \& Schultz (1997b) used a
steady--state model of ionization and thermal structure to calculate
early nebular--phase spectra for comparison with observed spectra
obtained hundreds of days after the explosion.  Although they favored
a {\sl sub}--Chandrasekhar mass ejection for SN~1991T, the mass was
higher than they favored for normal SNe~Ia (Liu, Jeffery, \& Schultz
1997a).

The argument that SN~1991T was super--Chandrasekhar depends, of
course, on our assumption that SN~1991T is at the same distance as
SNe~1981B and 1960F.  A Cepheid--based determination of the distance
to NGC 4527, although perhaps not easy for such a dusty, inclined
galaxy, is vital to check on this assumption.

This work has been supported by NSF grants AST 9417102, 9417242, and 9731450 
and NASA grant NAG 5--3505.

\clearpage

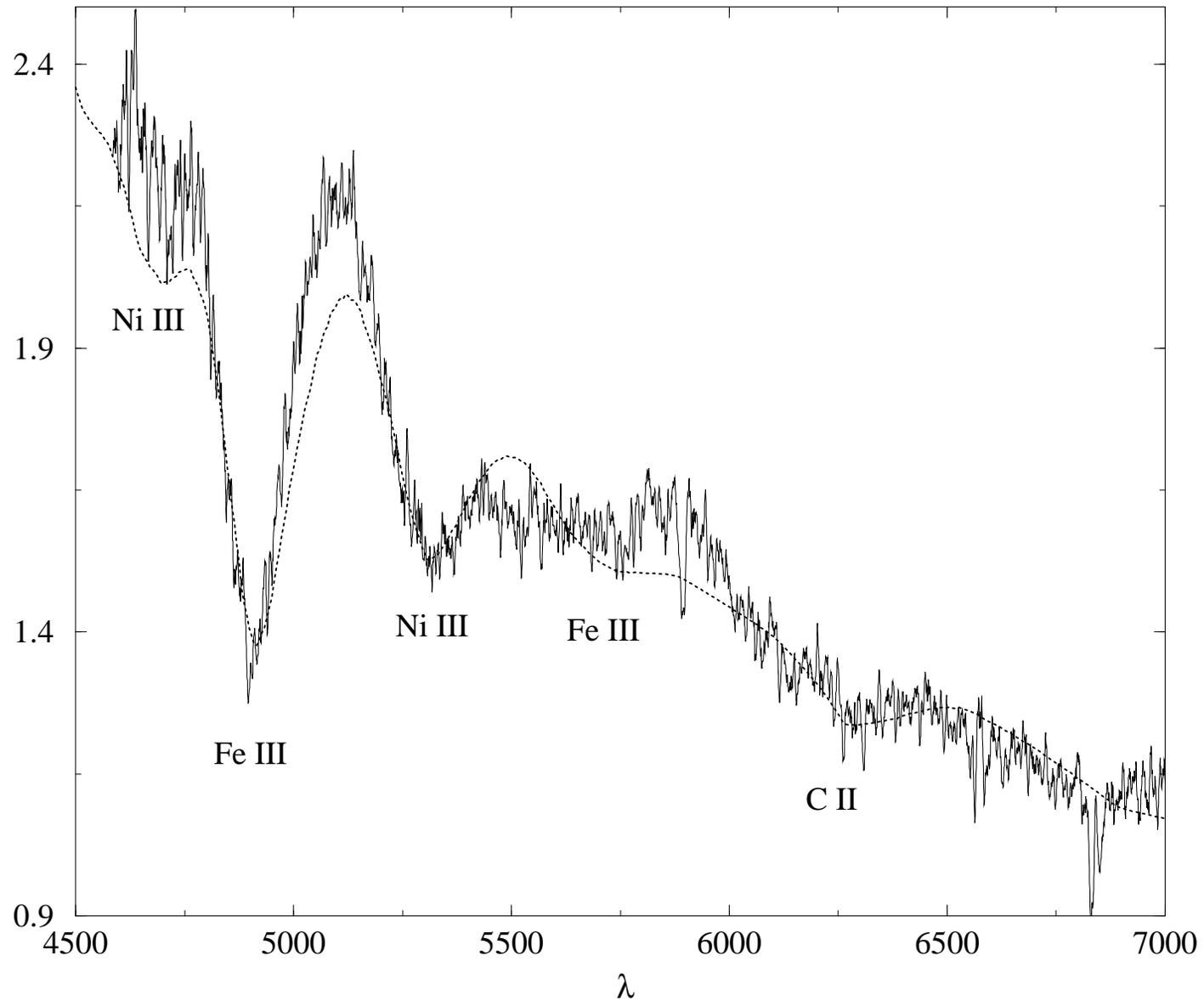
\begin{figure}

\figcaption{A spectrum of SN~1991T obtained at $-13$ days (Phillips
et~al. 1992) is compared to a synthetic spectrum that has
$v_{phot}=15,000$ \kms and $T_{exc}=13,000$ K. Ions responsible for
features in the synthetic spectrum are marked.
\label{fig1}} 
\end{figure}

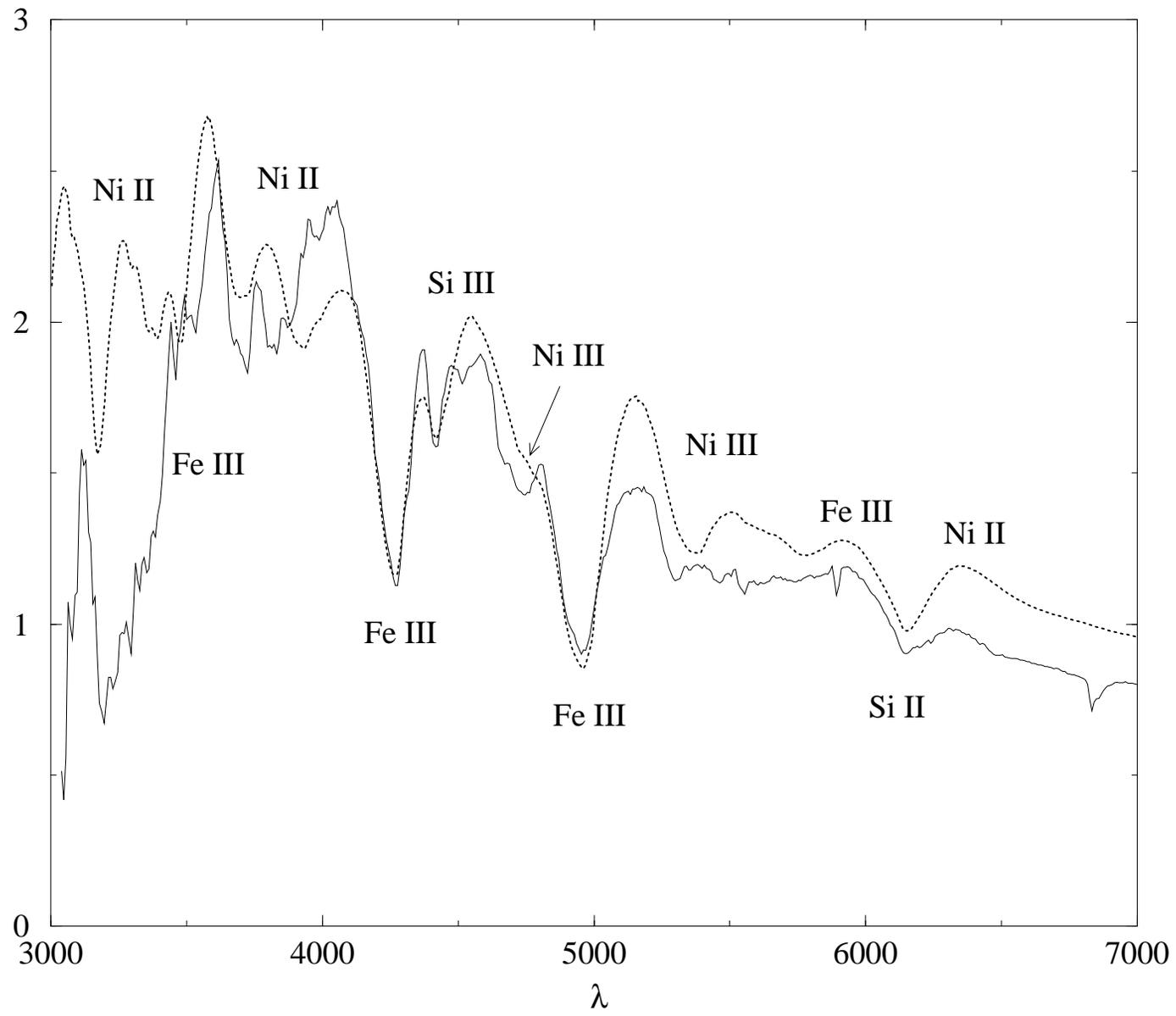
\begin{figure}

\figcaption{A spectrum of SN~1991T obtained at $-4$ days (Phillips
et~al. 1992) is compared to a synthetic spectrum that has
$v_{phot}=10,000$ \kms and $T_{exc}=13,000$ K.  Ions responsible for
features in the synthetic spectrum are marked.
\label{fig2}}
\end{figure}

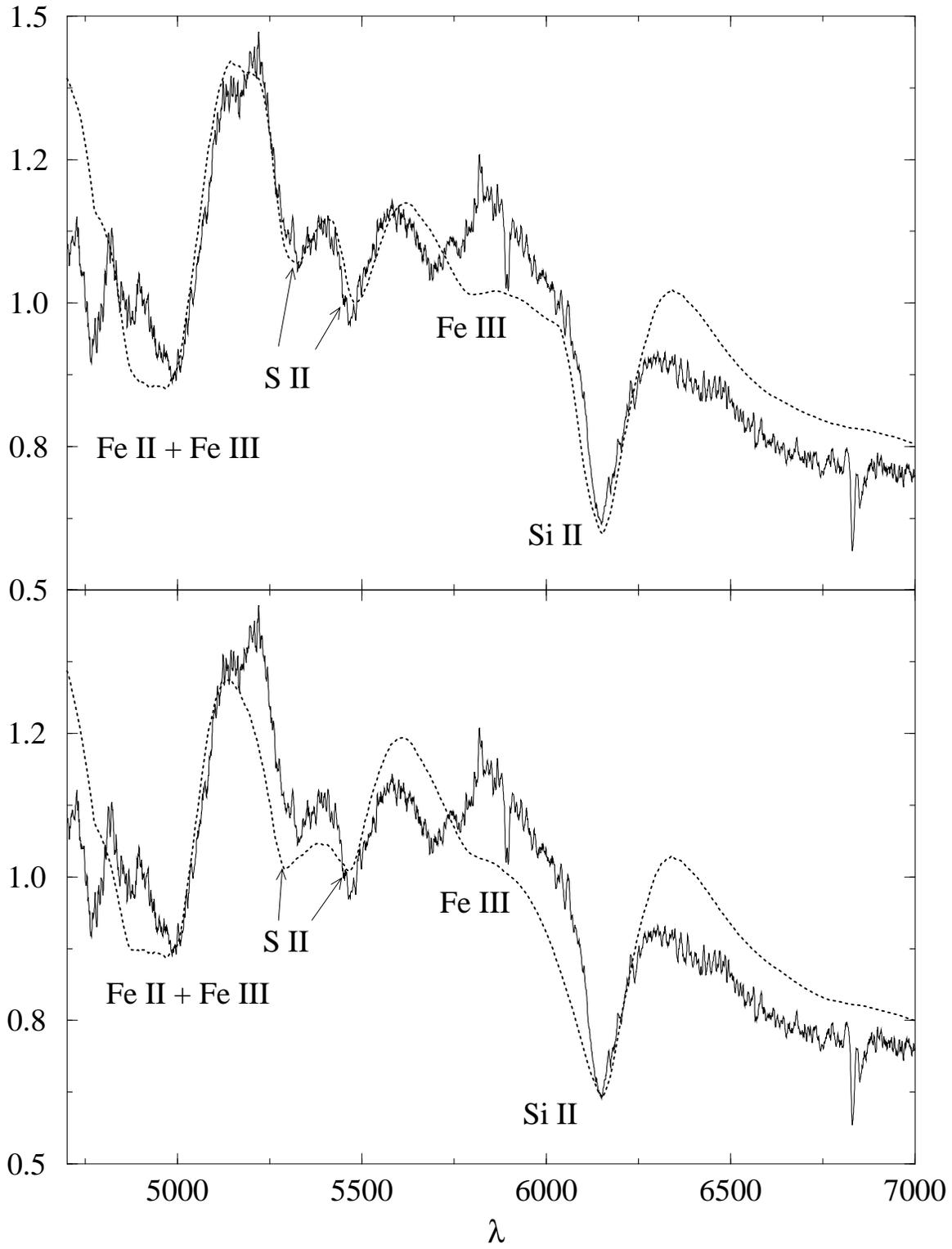
\begin{figure}

\figcaption{A spectrum of SN~1991T obtained at $+6$ days (Phillips
et~al. 1992) is compared to synthetic spectra that have
$v_{phot}=9500$ \kms and $T_{exc}=10,000$ K.  Ions responsible for
features in the synthetic spectrum are marked.  In the upper panel,
maximum velocities of 12,000 and 15,000 \kms have been imposed on S~II
and Si~II, respectively.  In the lower panel these limits have been
removed.
\label{fig3}}
\end{figure}

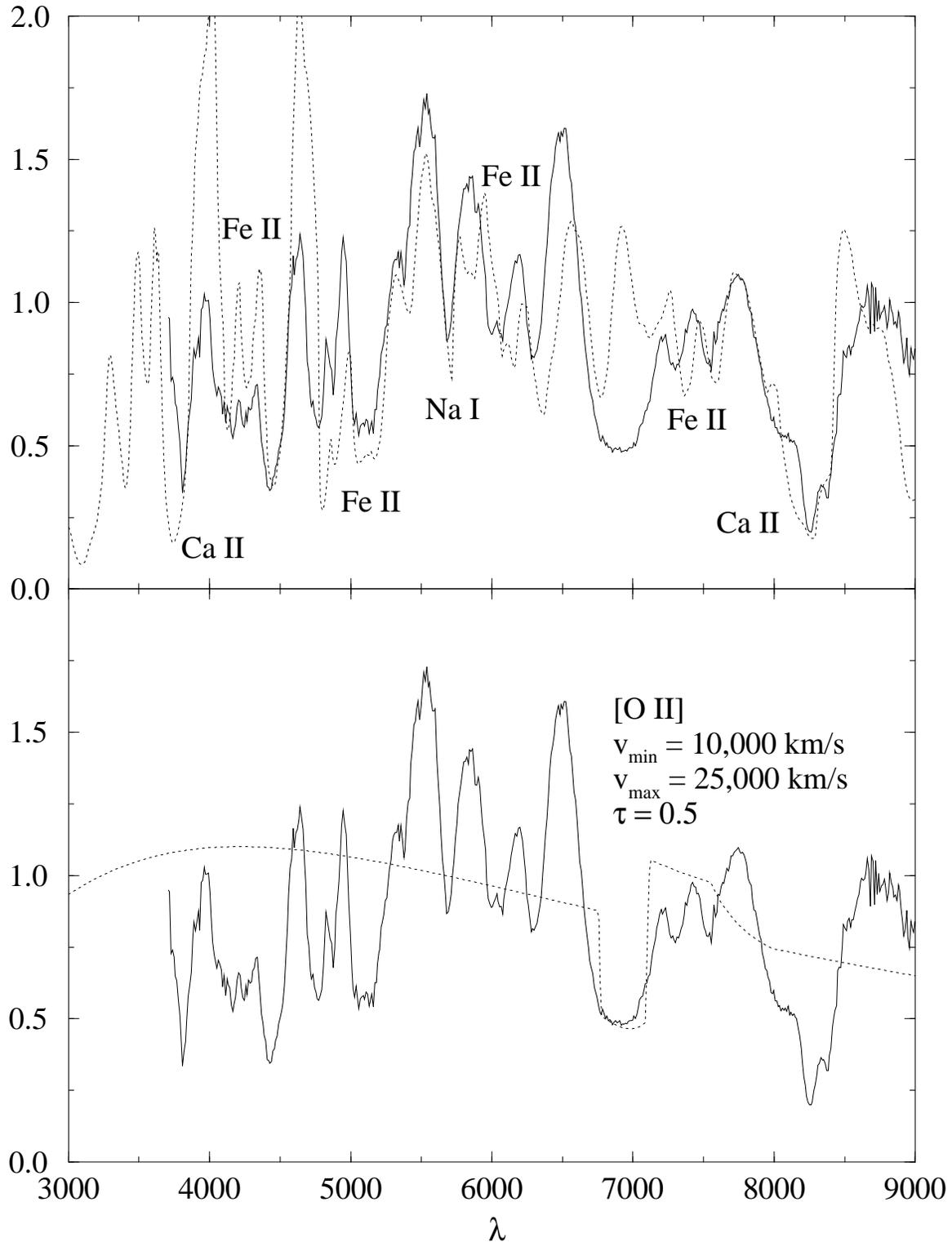
\begin{figure}

\figcaption{A spectrum of SN~1991T obtained at $+59$ days
(W. P. S. Meikle et~al., unpublished) is compared to synthetic
spectrum that have $v_{phot}=4000$ \kms and $T_{exc}=10,000$ K.  In
the synthetic spectrum of the upper panel, one feature is produced by
Na~I (detached at 9000 \kms), two features are produced by Ca~II, and
all others are produced by Fe~II (which has an abrupt factor of ten
decrease in its optical depths at 10,000 \kms).  In the synthetic
spectrum of the lower panel, the one feature is produced by [O~II].
\label{fig4}}
\end{figure}

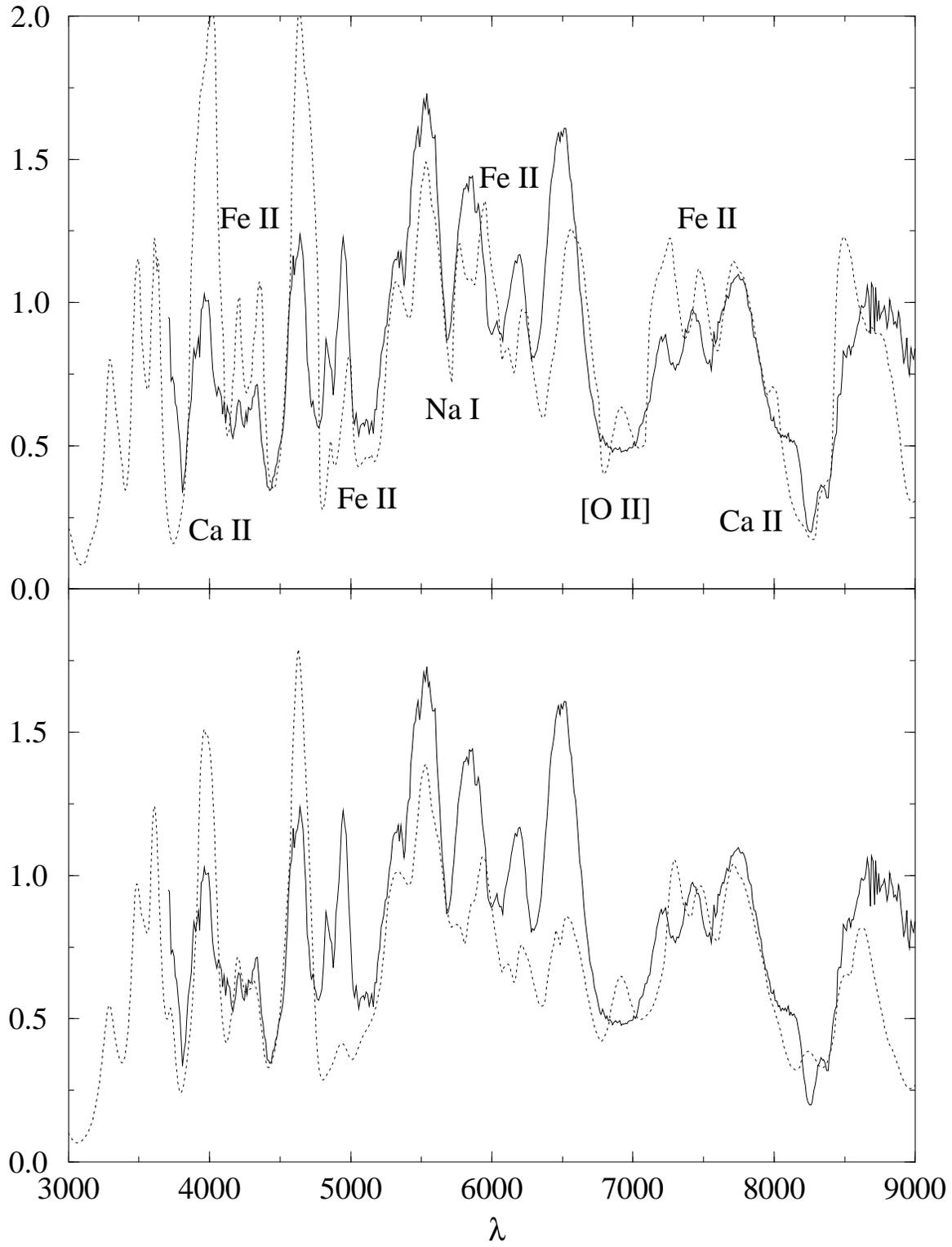
\begin{figure}

\figcaption{A spectrum of SN~1991T obtained at $+59$ days
(W. P. S. Meikle et~al., unpublished) is compared to synthetic
spectrum that have $v_{phot}=4000$ \kms and $T_{exc}=10,000$ K.  The
synthetic spectrum of the upper panel is like the one in the upper
panel of Figure~4, but now [O~II] is included.  In the synthetic
spectrum of the lower panel, the detachment of Na~I and Ca~II, 
the Fe~II optical
depth discontinuity, and the velocity limits on [O~II] have been
removed.
\label{fig5}}
\end{figure}

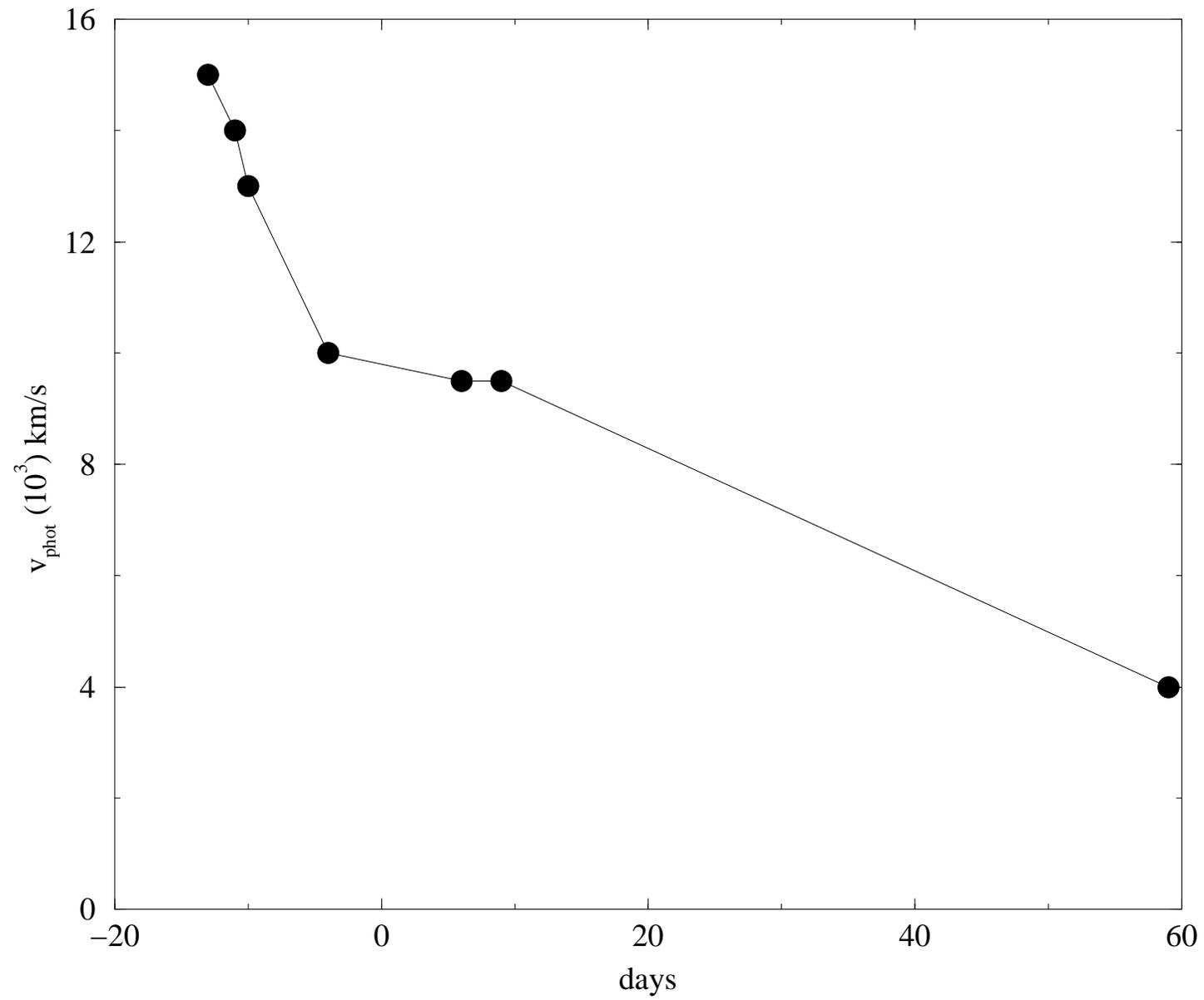
\begin{figure}

\figcaption{The velocity at the photosphere, used in the synthetic
spectra, is plotted against time.
\label{fig6}}
\end{figure}

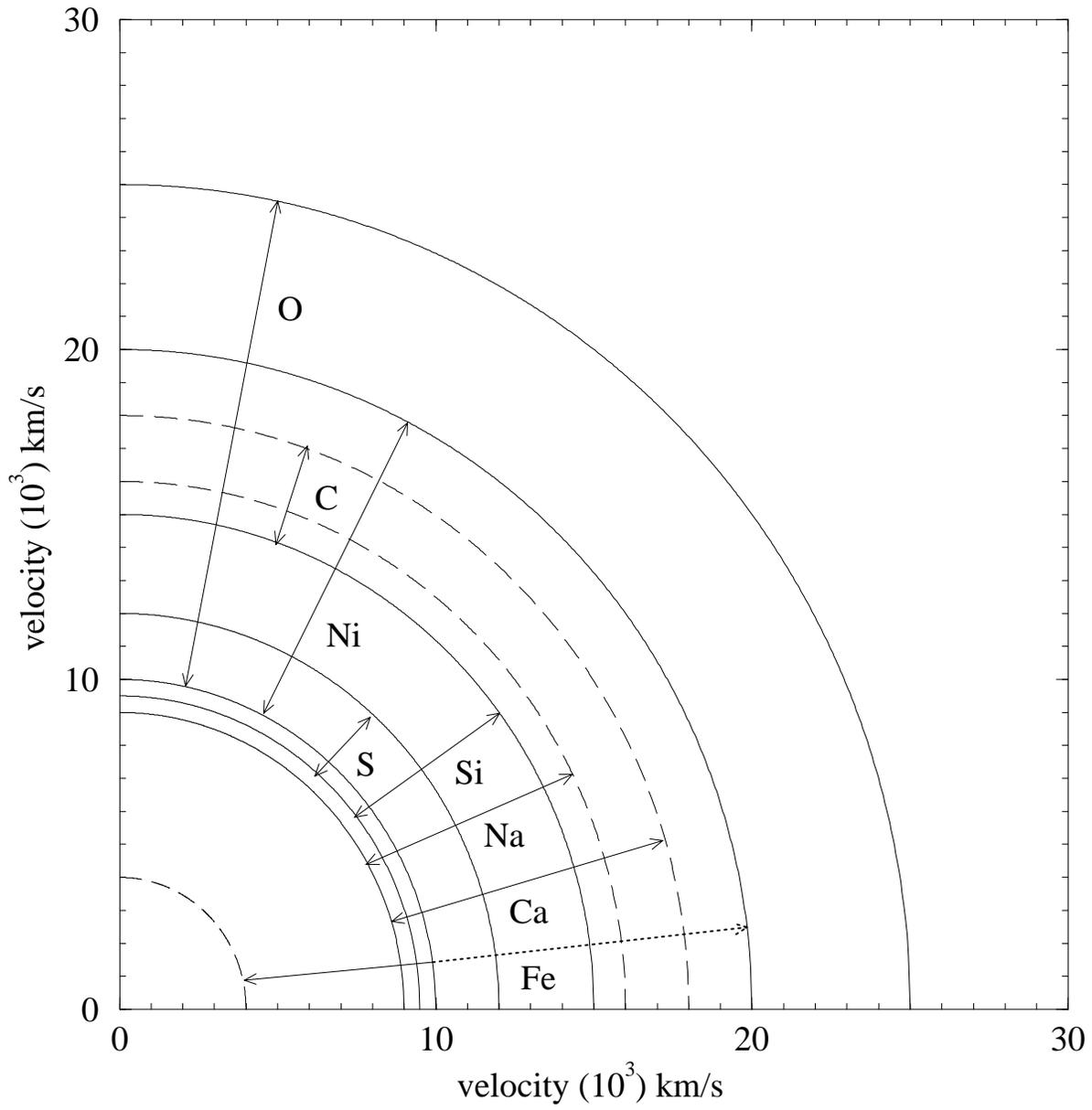
\begin{figure}

\figcaption{Inferred constraints on the composition structure.  Solid
arcs indicate maximum or minimum velocities that were imposed in the
spectrum calculations; for example, a maximum velocity of 15,000 \kms
was imposed on silicon.  Dashed arcs merely indicate
maximum or minimum velocities at which the element could be detected;
for example, calcium could be detected up to 18,000 \kms but there is
no evidence against its presence at higher velocities.  For Fe~II, the
transition from a solid line to a dotted one is at the velocity at
which the optical depth discontinuity was introduced.
\label{fig7}}
\end{figure}

\clearpage

\begin{deluxetable}{cccc}
\tablenum{1}
\tablewidth{0pc}
\tablecaption{Luminosities of Three Type Ia Supernovae}
\tablehead{
\colhead{} & \colhead{SN 1960F} & \colhead{SN 1981B} & \colhead{SN 1991T}
\\
\colhead{galaxy} & \colhead{NGC 4496} & \colhead{NGC 4536} & \colhead{NGC 4527}
}

\startdata
$B$         & $11.60 \pm 0.10$  &   $12.04 \pm 0.04$  &   $11.70 \pm 0.02$ \\
$V$         & $11.51 \pm 0.15$  &   $11.98 \pm 0.04$  &   $11.51 \pm 0.02$ \\
$E(B-V)$    & $0.0  \pm 0.00$  &    $0.10 \pm 0.05$  &    $0.20 \pm 0.10$ \\ 
$B^0$       & $11.60 \pm 0.10$  &   $11.63 \pm 0.20$  &   $10.88 \pm 0.41$ \\
$V^0$       & $11.51 \pm 0.15$  &   $11.67 \pm 0.16$  &   $10.89 \pm 0.31$ \\
$M_B^0$     & $-19.43 \pm 0.14$  &  $-19.47 \pm 0.24$  & $-20.19 \pm 0.43$ \\
$M_V^0$     & $-19.56 \pm 0.20$  &  $-19.43 \pm 0.21$  &  $-20.18 \pm 0.34$ \\
$M_{bol}-M_V$    & $0.10 \pm 0.10$  &    $0.10 \pm 0.10$  &   $-0.10 \pm 0.10$ \\
$M_{bol}$   & $-19.46 \pm 0.22$  &  $-19.33 \pm 0.23$  &  $-20.28 \pm 0.35$ \\
$L/10^{43} erg s^{-1}$ & $1.85 \pm 0.43$  &  $1.63 \pm 0.38$  &  $3.93 \pm 1.54$ \\
\enddata
\end{deluxetable}

\end{document}